\documentclass[nonacm]{acmart}
\usepackage{algorithm}
\usepackage{algpseudocode}

\usepackage{caption}
\usepackage{subcaption}
\usepackage{float}
\usepackage{amsmath}
\usepackage{verbatim}
\usepackage{ragged2e}

\DeclareTextFontCommand{\texttc}{\tcfamily}

\algnewcommand{\IIf}[1]{\State\algorithmicif\ #1\ \algorithmicthen}
\algnewcommand{\EndIIf}{\unskip\ \algorithmicend\ \algorithmicif}
\algnewcommand{\FFor}[1]{\State\algorithmicfor\ #1\ \algorithmicdo}
\algnewcommand{\EndFFor}{\unskip\ \algorithmicend\ \algorithmicfor}

\makeatletter
\newcommand*{\rom}[1]{\expandafter\@slowromancap\romannumeral #1@}
\makeatother
\makeatletter
\newcommand{\figcaption}[1]{\def\@captype{figure}\caption{#1}}
\newcommand{\tblcaption}[1]{\def\@captype{table}\caption{#1}}

\AtBeginDocument{%
  \providecommand\BibTeX{{%
    \normalfont B\kern-0.5em{\scshape i\kern-0.25em b}\kern-0.8em\TeX}}}

\acmConference[IUI 2022]{Helsinki, Finland '22: The 27th ACM Conference on Intelligent User Interfaces}{March 22--25, 2022}{Helsinki, Finland}
\acmBooktitle{Helsinki, Finland '22: ACM Conference on Intelligent User Interfaces,
 March 22--25, 2022, Helsinki, Finland}
\acmPrice{}
\acmISBN{978-1-4503-XXXX-X/18/06}

\begin{document}

\title{D-Graph: AI-Assisted Design Concept Exploration Graph}
\subtitle{}

\author{Shin Sano}
\authornote{Both authors contributed equally to this research.}
\email{ssano@nii.ac.jp}
\orcid{0002-1050-5008}
\author{Seiji Yamada}
\authornotemark[1]
\email{seiji@nii.ac.jp}
\affiliation{%
  \institution{Graduate University for Advanced Studies, SOKENDAI}
  \streetaddress{2 Chome-1-2 Hitotsubashi}
  \city{Chiyoda}
  \state{Tokyo}
  \country{Japan}
  \postcode{101-8430}
}

\renewcommand{\shortauthors}{Sano and Yamada}

\begin{abstract}
  We present an AI-assisted search tool, the ``Design Concept Exploration Graph” (``D-Graph”). It assists automotive designers in creating an original design-concept phrase, that is, a combination of two adjectives that conveys product aesthetics. D-Graph retrieves adjectives from a ConceptNet knowledge graph as nodes and visualizes them in a dynamically scalable 3D graph as users explore words. The retrieval algorithm helps in finding unique words by ruling out overused words on the basis of word frequency from a large text corpus and words that are too similar between the two in a combination using the cosine similarity from ConceptNet Numberbatch word embeddings. Our experiment with participants in the automotive design field that used both the proposed D-Graph and a baseline tool for design-concept-phrase creation tasks suggested a positive difference in participants' self-evaluation on the phrases they created, though not significant. Experts' evaluations on the phrases did not show significant differences. Negative correlations between the cosine similarity of the two words in a design-concept phrase and the experts' evaluation were significant. Our qualitative analysis suggested the directions for further development of the tool that should help users in adhering to the strategy of creating compound phrases supported by computational linguistic principles.

\end{abstract}



\maketitle

\section{Introduction}
Our goal is to develop and evaluate an intelligent UI that helps automotive concept designers to verbalize a original design concept with a phrase that expresses the aesthetics, mood, and emotional quality of a product. The contributions of this research are as follows. 1. We implement compound phrases using an adjective-adjective formula. 2. We implement algorithms that support users in creating a unique adjective-adjective phrase and terms that contrast with it to help create a character space.

\subsection{Aesthetic quality of design and its verbal representations}
The design in this study is primarily concerned with creating product aesthetics, defined as the characteristics that make up a product's appearance, including materials, proportion, color, ornamentation, shape, size, and reflectivity \cite{lawson2006designers}. In this section, we will discuss the relationship between product aesthetics and their verbal representations. Designers are in charge of creating the meanings and character attached to their designs and communicating them with other stakeholders in both visual and verbal modes \cite{chiu2012investigating, koch2019design,kita2018v8}. In the practices used in automotive design, designers exchange views on the form of a vehicle design as part of the design process. This requires them to communicate different shapes and features in words, such as ``slippery,'' ``exciting,'' ``fluid,'' ``tailored,'' and ``sheer.'' Such language used in car design studios describes particular forms or connotes a ``feeling.'' This language often tends to be ``idiosyncratic and atypical'' \cite{tovey1992intuitive}. Bouchard et al. \cite{bouchard2005nature} described these verbal expressions as ``intermediate representations (IR)'' and apparent especially in car design cases.

\subsection{Design concept phrase and character space}
We define a design concept phrase (``DCP'') as a combination of two adjectives that conveys product aesthetics. For example, ``kinetic warmth'' was created by the designers at Toyota’s North American design studio for Concept-i [50](Fig. 2-b). Note that the adjective ``warm'' was converted to a noun form. This unusually sounding DCP was iterated and communicated using the ``character space,'' which consists of two orthogonally crossed semantic differentials. A character space explains the design concepts in terms of how and by which attributes they differ and what already exists or what is to be avoided \cite{krippendorff2005semantic}. As illustrated by Tovey as ``idiosyncratic and atypical,'' a DCP should be non-cliché and use peculiar language to the extent that it calls for a design team to invent unique product aesthetics. While Han et al. \cite{han2021exploration} found that aesthetics have a positive relationship with the creativity of designs, there’s little computational support for verbally conceptualizing product aesthetics. To fill this gap, we present an AI-assisted search tool, the ``Design Concept Exploration Graph'' (``D-Graph''), which assists automotive designers in creating a novel design-concept phrase.

\begin{figure}[t]
  \centering
  \includegraphics[height=4.5cm]{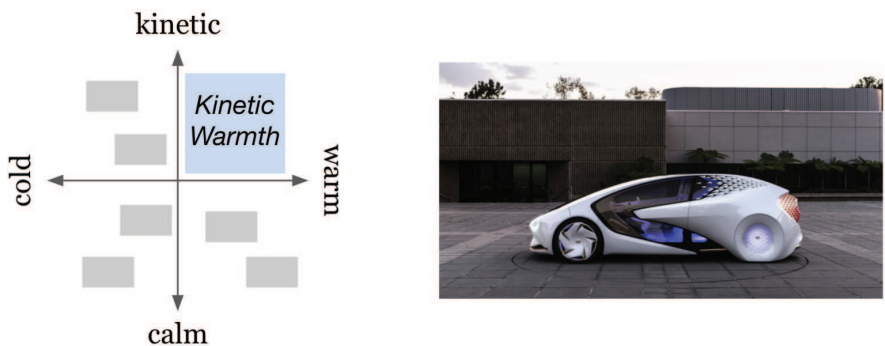}
  \caption{(a) left:The Character space for ``kinetic qarmth'' (b) right: the design derived from ``kinetic warmth''(© 2021 Toyota Motor Sales, U.S.A., Inc.).}
  \Description{}
\end{figure}

\section{Related Work}

\subsection{Computational mind-mapping tools}
While our D-Graph uses a visualized knowledge graph, other computational mind-mapping tools may seemingly look similar. \textit{Spinneret}\cite{bae2020spinneret} uses the ConceptNet knowledge graph and provides idea suggestions as nodes on the graph. It uses random walks with breadth-first and depth-first search biases to explore ideas and provides ``suggestions'' for the user to add to a mind map. Its goal is to support divergent thinking, so the algorithm attempts to pursue depth for concepts in the graph and to overcome the ``design fixation'' \cite{jansson1991design} problem. \textit{Mini-Map} \cite{chen2019mini} also uses ConceptNet, and it gamifies the process of mind-mapping, where a user collaboratively creates a mind map with an intelligent agent. The authors measured the diversification of ideas in a graph by using dimensionality reduction for the semantic distance between the ideas in mind-maps. Our approach is different as our goal is to assist users in examining the nuances of words within a particular semantic space, and the distance is only incorporated between the two words in the DCP.

\subsection{Combinational creativity}

Combinational creativity produces new ideas by combining existing ideas in unfamiliar ways \cite{boden2001creativity}. A number of studies have identified the effect of language as a tool for generating creative design concepts \cite{han2020computational, Georgiev2010d,Chiu2012g}. Nagai et al. \cite{nagai2009concept}, in a collaboration with designers, suggested that a new concept can be created by synthesizing two ideas in noun-noun combination phrases. Chiu and Shu \cite{Chiu2012g} used oppositely and similarly related pairs of words as stimuli for design-concept generation and observed that oppositely related verb-verb combination stimuli could increase concept creativity. Combinator \cite{han2016combinator} imitates the way the human brain achieves combinational creativity and suggests combined ideas in both visual and textual representations. However, in these pieces of research, combinational creativity focused either on combined product categories, derived from noun-noun combinations (e.g., ``desk + elevator'' or ``pen + ruler'') or functional features derived from verb-verb combinations (e.g., ``fill + insert''), and it did not address product aesthetics. In this paper, we aim to generate combinational creativity as adjective-adjective (or noun, e.g., ``kinetic + warm(th)'') forms. To check this, we conducted a preliminary study comparing the effects between noun-noun, verb-verb, and adjective-adjective (adjective-noun) forms of phrase.

\subsection{Utilizing lexico-semantic features in concept generation}
In relation to the method used in this study, we focus on two key features: word frequency and cosine distance between words. Setchi et al. \cite{Setchi2011a, setchi2011semantic} used \textit{term frequency-inverse document frequency} (TF-IDF) measurement to identify the most meaningful words around an image and added new tags to it with terms retrieved by semantic expansion from the original terms. The method demonstrated that a term with a low document frequency in a corpus could support richer inspiration and creativity for automotive designers. 
Han et al. \cite{han2019three, han2018conceptual} analyzed the conceptual distances between two ideas expressed in words (base and additive) that were incorporated in each of 200 award-winner product-design descriptions and measured the correlations between the distance of two words and expert designers' objective evaluations in ranking creativity. The result showed that good design concepts fell into a certain range of distance between two ideas. Also, a comparative study on using different language models to measure the conceptual distances between two ideas suggested that ConceptNet best agrees with human experts' judgement on concept distances \cite{han2020computational}. In D-Graph, we also use the frequency and cosine distance. What is unique about our method is that it uses them in a sequence to control the quality of the combinational adjectives as well as searches for contrasting concepts (antonyms) for each to help the user explain design concepts in terms of how and by which attributes they differ.

\section{Method}
\subsection{Preliminary study}
We decided to exclude noun-noun combinations to avoid words like ``desk-elevator,'' which suggests a ``mashup'' of two objects rather than collectively expressed product aesthetics. Meanwhile, adjective-noun combinations can be used as long as the second word is a noun form of an adjective so that it forms a compound adjective in a restrictive sequence - a sequence in which an adjective restricts another adjective that follows \cite{sopher1962sequence}. As in the example of ``kinetic warmth'' introduced earlier, conversion to a noun can be done afterwards to abbreviate the noun to be modified. Yet, we needed to know whether adjective phrase would work better over others, so we conducted a preliminary experiment comparing phrases with different parts-of-speech (PoS) so that we could be confident that adjective-adjective combinations are suitable for communicating product aesthetics over other PoS. We used Survey Monkey to recruit 55 participants from the U.S. whose job function is ``Arts and Design.'' Six sample phrases were randomly generated for each of four different combinations of PoS: adjective-adjective (AA), adjective-noun (AN), noun-noun (NN), and verb-verb (VV). The participants rated 24 randomly generated two-word combinational phrases in randomized order on a 7-point Likert scale on the basis of the degree to which they agreed with the statement ``I can imagine the product aesthetics, characters, mood, and emotional quality that are conveyed by the product design.'' An ANOVA-Tukey test identified homogeneous subsets as [AA/AN/NN] and [VV/NN], where the mean score of the former was significantly higher $(p =0.05)$. 

    The further details of the individual comparisons are as follows.  There was a significant difference
    \begin{math}
      (p =.038) 
    \end{math}
in mean scores between AA
    \begin{math}
      (3.890, \sigma =1.121) 
    \end{math}
and VV 
    \begin{math}
      (3.300, \sigma =1.205) 
    \end{math}
, a significant difference
    \begin{math}
      (p =.040) 
    \end{math}
between AN
    \begin{math} 
    (3.885, \sigma =1.025) 
    \end{math}
and VV
, no significant difference
    \begin{math}
      (p =.581) 
    \end{math}
between AA and NN
    \begin{math}
      (3.612, \sigma =1.232) 
    \end{math}
, no significant difference
    \begin{math}
      (p =.599) 
    \end{math}
    in between AN and NN
, no significant difference
    \begin{math}
      (p =.485) 
    \end{math}
between NN and VV
, and no significant difference
    \begin{math}
      (p =1.000) 
    \end{math}
between AA and AN. We randomly extracted the words for all combinations from a corpus we created with English automotive design articles containing about 1.4 million words on Sketch Engine\cite{kilgarriff2004itri}. The following filtration was performed equally to all PoS combinations. First, we removed non-English terms and proper nouns, then removed words, whose relative frequency per a million tokens was less than 1.0, compared to enTenTen15 \cite{jakubivcek2013tenten}, a large web text corpus containing about 1.6 billion words. This makes the list of all words, regardless of PoS, more suitable for design concept expressions that participants can readily recognize and make a judgement.

\subsection{System and UI}
\begin{figure}[h]
  \centering
  \includegraphics[width=\linewidth]{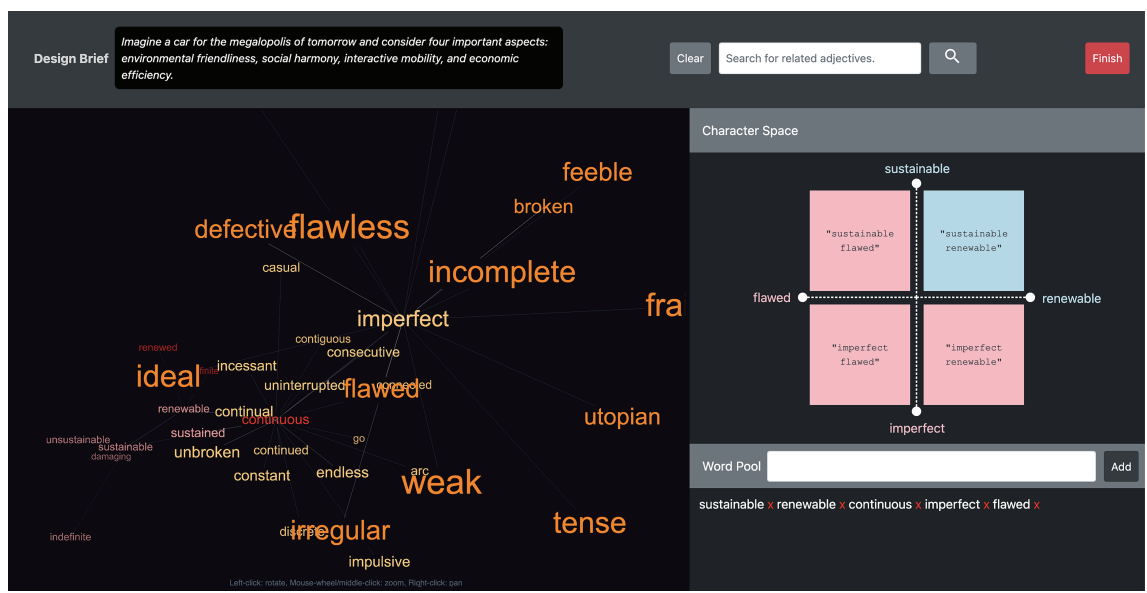}
  \caption{D-Graph (experiment) web UI with 2-A participant's character space and word pool replicated for analysis purpose}
  \Description{}
 \end{figure}
 
 \begin{figure}[h]
  \centering
  \includegraphics[width=\linewidth]{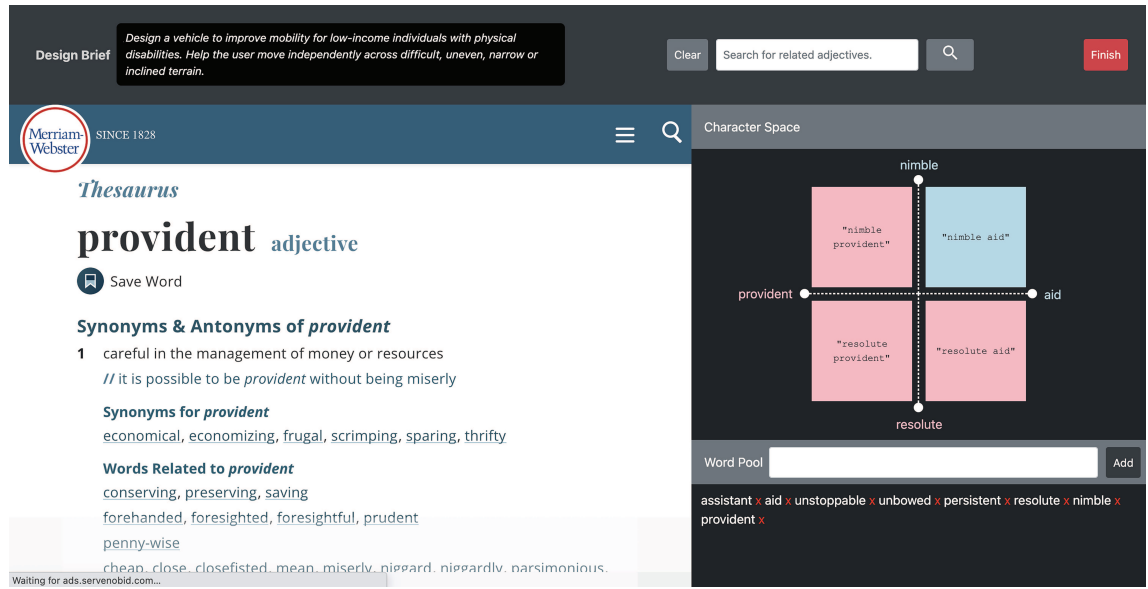}
  \caption{Baseline web tool without D-Graph UI and search algorithm. We used the Merriam Webster online thesaurus instead. Replicated screen from 5-B participant's character space and word pool for analysis purpose.}
  \Description{nimble aid/sustainable renewable}
 \end{figure}

The algorithm of D-Graph searches for and filters adjectives that are related to users' queries by using a ConceptNet knowledge graph \cite{speer2017conceptnet}. ConceptNet is a multilingual knowledge graph that connects words and phrases with labeled edges, and it has the most extensive relations of edges over other commonly available knowledge graphs. The top section of the web UI has a design brief and a search window. The large space below them is allocated to a ``playground,'' in which graphs of explored words are visualized in 3D hub-and-spoke style. Users can examine the words at different granularities by zooming in/out, rotating, and panning the graph. Every time the user expands the search by clicking words, new clusters are shown in different colors so that users can visually track back to previous explorations. A word definition can be seen by mousing over the words. The lower-right section is a ``word pool'' where users can store favorite words as candidates for design concept phrases. Every time the user puts a query in the search window, clicks on a word in the playground, or drags \& drops words from the playground to the word pool, those words are stored in the word pool. Finally, the right-middle section is the character space (CS), which consists of two semantic differentials orthogonally crossing each other. The CS is completed when all four ends on the axes are defined as word 1 ($w_1$), word 2 ($w_2$), word 3 ($w_3$), and word 4 ($w_4$). The words on the CS can be set by dragging \& dropping words either from the word pool or directly from the graphs in the playground. The upper-right quadrant, represented by the combination of $w_1$ and $w_2$, is the target design-concept phrase. All the other quadrants, represented by $w_2$/$w_3$, $w_3$/$w_4$, and $w_4$/$w_1$, are contrasting concepts to be used by the users to explain the target design concepts in comparison with opposing concepts.

\subsection{System configuration}
D-Graph consists of a front-end web application written in JavaScript, HTML, and CSS and a back-end web server written in Python with a MySQL database hosted on PythonAnywhere \cite{pythonanywherepythonanywhere}. D3.js \cite{zhu2013data} is used to visualize the graph data structure. MySQL database holds the data of word embeddings from ConceptNet Numberbatch \cite{speer2017conceptnet} and word frequencies in enTenTen15 \cite{jakubivcek2013tenten}, an English web corpus retrieved via Sketch Engine \cite{kilgarriff2014sketch}.

\subsection{Search and filter algorithms}
D-Graph directs users to set words on the CS in a predetermined order ($w_1$, $w_2$, $w_3$, then $w_4$). This strategy mandates that users first establish the target design concept, represented by the upper-right quadrant (blue square shown in Fig. 2) defined by $w_1$ and $w_2$, followed by the contrasting concepts (all other quadrants shown in pink in Fig. 2) that can be used in a ``not this, but this" style of explanation. D-Graph has two types of search and filter algorithms, \small{SEARCH\_FOR\_RELATED\_WORDS} \normalsize(Algorithm 1) and \small{SEARCH\_FOR\_ANTONYMS} \normalsize (Algorithm 2). The former gets a new word $w'$ from all nodes inside the edge $e$, whose relation to the query $w$ is NOT ``Antonym." This means it uses all other 33 relations in ConceptNet in order to maximize the number of words retrieved. Then, each new word $w'$ from the nodes will be filtered in terms of both the relative word frequency($Freq$) of $w'$ (number of occurrences per one million tokens in the enTenTen15 English corpus) and the cosine similarity ($cosSim$) between the query word $w$ and the new word $w'$, calculated with ConceptNet Numberbatch word embeddings. After several tests, we currently set the threshold at $(.05 \leq |cosSim| \leq .5)$ and $(1 \leq Freq \leq 50)$. The latter (Algorithm 2) first gets related words with the former algorithm; then, for each new word $w'$, it gets all the nodes in the edge whose relation to $w'$ is ``Antonym." All the results are set in the D3.js as labels of the start node and end node and the link between them to render the graph.

\begin{algorithm}
    \caption{Search for Related Words}
\begin{algorithmic}[t]
        \Function {search\_for\_related\_words}{$w$}
            \State {Initialize $results$ = [] } 
            \For {each edge $e$ for word $w$ in ConceptNet} 
                \If {relation of $e$ is not ``Antonym''}
                    \State {$w'$ := \texttt{get\_other\_word}($e$, $w$)}
                    \State {$cosSim$ := \texttt{get\_cosine\_similarity}($w$, $w'$) from ConceptNetNumberBatch}
                    \State {$Freq$ := \texttt{get\_relative\_frequency}($w'$) from SketchEngine}
                    \IIf {$(0.05 \leq |cosSim| \leq 0.5) \bigwedge (1 \leq Freq \leq 50)$} {$results$.append($w'$)}
                    \EndIIf
                \EndIf
            \EndFor
            \State \Return $results$
        \EndFunction
    \end{algorithmic}
\end{algorithm}
\begin{algorithm}
    \caption{Search for Antonyms}
\begin{algorithmic}[b]    
        \Function {search\_for\_antonyms}{$w$}
            \State {Initialize $results$ = []. } 
            \State {Initialize $related\_words$ = \small{SEARCH\_FOR\_RELATED\_WORDS}($w$)} 
            \For {each word $w'$ in $related\_words$} 
                \For {each edge $e$ for word $w'$ in ConceptNet} 
                    \If {relation of $e$ is ``Antonym''}
                        \State {$s$ := \texttt{get\_start}($e$)}                    
                        \IIf {$s$ is not in $results$} {$results$.append($s$)}
                        \EndIIf
                        \State {$d$ := \texttt{get\_end}($e$)}                    
                        \IIf {$d$ is not in $results$} {$results$.append($d$)}
                        \EndIIf
                    \EndIf
                \EndFor
            \EndFor
            \State \Return $results$
        \EndFunction
    \end{algorithmic}
\end{algorithm}

\section{Experiment Design}

\subsection{Participants and independent variables}
Eleven undergraduate/graduate students (10 M, 1 F), who specialize in automotive design in an industrial design department were, recruited. Unfortunately, the only female participant's responses became invalid due to an inadequate task environment, which left us 10 all-male valid participants, with a mean age of 25.1 years ($sd =4.01$). The independent variables were two different search tools, D-Graph (Fig. 2) and the baseline tool (Fig. 3) using the Merriam-Webster online thesaurus \cite{webster2014merriam}. As mentioned above, D-Graph consists of search and filter algorithms with a 3D graph visualizer. The UI of the search window, word pool, and CS [of the two tools?] were the same, except for the word set order algorithm in D-Graph. The participants were asked to perform the same task twice with the baseline and experimental tools in a counterbalanced order (Table 1). They participated in the experiment online using Playbook UX, a usability testing platform that enables screen and ``think-aloud'' audio recordings. Each participant was given a video instruction explaining the goal of the task and how to use each tool before the task. They were also given a practice time window (2-3 min.) to get familiar with how to operate the tools between the video instruction and actual experiment.

\subsection{Stimuli and tasks}
All participants were provided with two different design briefs (Table 2). A design brief is a written description of a project that requires some form of design, containing a project overview, its objectives, tasks, target audience, and expected outcomes \cite{phillips2004creating,koronis2018impact}. 
The two briefs were also given to the participants in a counterbalanced order (Table 1). After reading the brief, the participants were prompted to start the task. First, they were asked to find a combination of two words that forms a DCP by determining $w_1$ and $w_2$; then, they were asked to find the opposing concept to each of $w_1$ and $w_2$ to generate the CS \cite{krippendorff2005semantic}. The session was concluded when the user was satisfied with the design concept phrase in terms of $w_1$ and $w_2$ and comfortable explaining it in contrast to the other three quadrants.

\begin{figure}
\begin{minipage}[t]{.45\textwidth}
\justify
\def\@captype{table}
  \caption{Counterbalanced task arrangement}
  \label{tab:commands}
  \begin{tabular}{ccc}
    \toprule
    Task &Group 1 ($N=5$) & Group 2 ($N=5$)\\
    \midrule
     1 &Exp.+ D.brief A &Cont + D.brief B \\
     2 &Cont+ D.brief B &Exp. + D.brief A \\
    \bottomrule
  \end{tabular}
\end{minipage}
\begin{minipage}[t]{.45\textwidth}
    \centering
    \def\@captype{table}
        \caption{Two variations of design brief}
        \begin{tabular}{ll} \toprule
            \small A\cite{setchi2011semantic}:&
            \begin{tabular}{l}
                \small Imagine a car for the megalopolis of \\
                \small tomorrow and consider four important aspects:\\
                \small environmental friendliness, social harmony, \\
                \small interactive mobility, and economic efficiency.\\
            \end{tabular}\\ \midrule
                \small B\cite{koronis2020best}:&
            \begin{tabular}{l}
                \small Design a vehicle to improve mobility for \\
                \small low-income individuals with physical disabilities.\\ \small Help the user move independently across difficult,\\ \small uneven, narrow or inclined terrain. (modified)
                \\
            \end{tabular}\\ \bottomrule
        \end{tabular}
\end{minipage}
\end{figure}

\subsection{Quantitative measurement - dependent variables}
\subsubsection{Subjective evaluation}
A post-task questionnaire with self-reported evaluations was administered using a 7-point Likert scale for three measurements: ``originality'' of the DCP, ``relevancy'' of the phrase to the design brief, and the ``explainability'' of the DCP.  The participants were asked to write a short explanation of the DCP (upper-right quadrant of the CS), in contrast to the ideas expressed in the other quadrants.  This was measured by a 7-point Likert scale on how comfortable they were in explaining the DCP.

\subsubsection{Objective evaluation}
Since this study is highly domain specific, the evaluation should also be done by the consensus of expert raters\cite{shah2003metrics}. Two former executives in the automotive design field who have led global design teams were recruited. They were asked to rate each design concept phrase in a randomized order while being blinded about which of the tools were used to generate each DCP. 7-point Likert scales were used for the ``originality'' of the DCP, ``relevancy'' of the DCP to the design brief, and the ``imageability'' of the DCP for the degree to which they could imagine the design of the vehicle described with this design concept phrase. The inter-rater reliability using Cohen's kappa coefficient on the metrics above were $k=.352,(p=.011), k=.159 (p=.184), k=.215 (p=.065)$, respectively.

\subsubsection{Computational metrics}
The relative word frequency ($Freq$) of both $w_1$ and $w_2$ for each DCP as well as the cosine similarity ($cosSim$) between them were calculated post-hoc. The duration of the task and the word count in the ``word pool,'' which indicates how many words the participant interacted with in the task, were also retrieved. In addition, we measured the Pearson's correlations between the experts' ratings on ``originality'' and the linguistic metrics above.  We further analyzed how selected participants interacted with the words using spacial mapping based on the word embedding.

\subsection{Qualitative data}
All the DCPs and two other words on the CS (Table 6), as well as the written explanations, were obtained. We also had the screen recordings of all sessions as well as the think-aloud protocols. The participants were asked to verbalize their thoughts during the session. The think-aloud protocols were hand-coded and summarized.  Video recording was analyzed to acquire insights on how the participants actually utilized the tool overall, including the participants strategy, mental model, usability issues, and how they should be addressed in the future development.

\section{Results}
\subsection{Quantitative measurement} 
\subsubsection{DCPs evaluation by participants results}
No significant differences were found between the D-Graph and the baseline tool in participants' subjective evaluations on DCPs (Table 3).

\subsubsection{DCPs evaluation by experts results} No significant differences were found between the D-Graph and the baseline tool in experts' objective evaluations on DCPs.

\subsubsection{Computational Metrics}
No significant differences were found on the task duration and word count in the word pool (Table 4).  Table 5 shows all the DCPs with the participant ID with the type of design brief (A/B) , the tool used, the word count (WC) in the word pool, the relative word frequency($Freq$) of both $w_1$ and $w_2$, and the mean between the two in each DCP, as well as the cosine similarity ($cosSim$) between them.  Some words did not return a value due to the unavailability of them in the word embeddings or in the corpus. There were no significant differences $(p=.218)$ in mean $cosSim$ between the D-Graph $(.246, \sigma =.195)$ and the baseline tool $(.149, \sigma =.124)$ and no significant difference $(p=.154)$ in the mean of mean-$Freq$ between the experiment tool $(26.42, \sigma =20.76)$ and the baseline tool $(13.73, \sigma =15.61)$.

Interestingly, there was a moderate negative Pearson correlations $(-.632, p=.004)$ between $cosSim$ and the experts' objective evaluation on ``originality,'' and a high negative Pearson correlations $(-.758, p\leq.001)$ between mean-$Freq$ and the experts' objective evaluation on ``originality.''(Figure.4) No such significant correlations were found between the same linguistic metrics above and the participant's subjective evaluations.  We also examined how the participants interacted with the words, got the words in the word pool, and decided $w_1-w_4$ along with the computational linguistics metrics. We will present summaries of four cases in this paper. Fig. 6 shows the participant's exploration process where circled numbers are the sequence of the interactions. The words are mapped according to the the ConceptNet Numberbatch word embeddings, whose dimensionality is reduced by principal components analysis (PCA). The blue areas show searches for DCPs for $w_1$ and $w_2$, and the pink areas show searches for antonyms for $w_3$ and $w_4$, $w_3$.

    {\bfseries Case 1-B:``protean companionable''} (Fig.6-(a)) had a $cosSim$ value of $0.063$, the smallest $meanFreq$ was $0.13$, created with the baseline tool responding to the design brief B. The experts ``originality'' score marked $6.5$. The number of the words in the word pool was $20$ and the task duration was 12 minutes and 59 seconds. He started with `versatile,' and typed the next three words without clicking on the words in the thesaurus, then found `protean' in the thesaurus.  He then restarted the search by typing `companionable,' then found the next three words in the thesaurus. Then he put `affordable', mentioning \textit{``...now I come back to a kind of basic''} and found the next four words in the thesaurus. At this point, he had already decided to put $w_1$(``protean'') and $w_2$ (``companionable'') for the DCP, and started exploring antonyms. He found `limited' as an antonym of `protean' and found the next four words in the thesaurus. He relied on his own judgement of the first four queries and started to interact with the thesaurus more except for two ``fresh starts'' on `companionable' and `affordable'.  His own ratings for ``originality'' and ``relevancy'' were $5$ and $7$. 
   
    {\bfseries Case 1-A:``cognizant inclusive''} (Fig.6-(b)) was made by the same participant above using the D-Graph responding to the design brief B. It has a $cosSim$ value of $0.105$, the $meanFreq was (10.37)$. The experts ``originality'' score marked $5.0$. The number of the words in the word pool was $23$ and the task duration was 14 minutes and 58 seconds. He typed the first word `amiable' and used manual search window, instead of clicking the words on the graph, until he found the sixth word, `visionary' on the D-Graph.  During that time, he opened a new tab on the browser and used an online thesaurus to find the adjective form of `utopia' as it was denied by the system because `utopia' was not an adjective.  Then the words he explored aimed to express ``being aware of the social issues''.  He also noted,``Desirable for sure, but that's given.'' By the time he stored the 16th word (`citywide') in the word pool he decided to use `cognizant' and `inclusive' for the DCP.  Putting `cognizant' on $w_1$ triggered showing the candidates for $w_3$, he found `oblivious', and dropped it to $w_3$ as he noted, ``that's a good word.'' It triggered showing the candidates for $w_4$, but it showed only four words, including the root node(`inclusive', `micro', `exclusionary', and `exclusive.' He tried `micro' anyway, but he did not find anything he liked.  So he went back to `inclusive' by clicking it, and clicked on `exclusive' that gave him eighteen new words.  After examining all words there he dragged and dropped 'selective' to $w_4$. His own ratings for ``originality'' and ``relevancy'' were $4$ and $7$. 
    
    {\bfseries Case 2-A: ``sustainable renewable''} (Fig.6-(c))
    was made using the D-Graph responding to the design brief A. It has a $cosSim$ value of $0.572$, the $meanFreq was (66.74)$ The experts ``originality'' score marked $1.5$. The number of the words in the word pool was $5$ and the task duration was 1 minutes and 23 seconds. Given the small numbers of words in the word pool the participants interacted for the next two cases, we will also present the number of words added when a word was clicked in the playground in parentheses, retrieved from the JavaScript console in a replicated session. He first typed `sustainable'($6$) in the search window after reading the design brief that showed six candidates for $w_1$. Then he clicked `renewable' ($5$), then `continuous' ($17$), `imperfect' ($14$), and, finally, `flawed' ($1$). Once `continuous' was clicked, it took him only 14 seconds to click the latter three words. He dragged and dropped `sustainable' for $w_1$, followed by `renewable', `imperfect'' and `flawed' for $w_2$, $w_3$, and $w_4$, respectively.  In this session we observed the participant put all the words at once in $w_1$ through $w_4$, where the new search for distant word for $w_2$, as well as antonym searches triggered by setting $w_2$ and $w_3$, were not utilized. Since he used the first two words including his own query, the $cosSim$ value between two words were among the largest.  His own ratings for ``originality'' and ``relevancy'' were $4$ and $7$. 
    
    {\bfseries Case 7-B: ``economical efficient''} (Fig.6-(d)) was made using the D-Graph responding to the design brief B. It has a $cosSim$ value of $0.551$, the $meanFreq was (31.64)$ The experts ``originality'' score was $2.5$. The number of the words in the word pool was $6$ and the task duration was 7 minutes and 1 second. After reading the design brief, he typed `economical'($5$) in the search window that showed five words. After clicking on `efficient'($18$)	and `capable'($25$) he spent a minute and forty seconds to rotate the graph and mouse-overed several words to see the definitions, click `efficient' and `capable' twice each, then finally cleared the playground and typed `economical' again, followed by clicking `efficient.' Then he clicked `futile', but it was apparently accidental as he deleted `futile' soon and cleaned up the playground again.  He typed and clicked `efficient' , `capable', and `capable' for the third time.  Before clicking the next one, `resourceful'($5$), he carefully examined the definition of `competent', `thorough' and `resourceful.' Then he spent twenty seconds to see the definition of `ingenious', pause another ten seconds to click `ingenious'($5$), followed by `natural'($5$) in fifteen seconds. He further spent fifty-two seconds to rotate the graph, clicked `capable' and `resourceful' again, then put `economical', `efficient',  `capable', and  `resourceful' for $w_1$, $w_2$, $w_3$, and $w_4$, respectively. His own ratings for ``originality'' and ``relevancy'' were $6$ and $7$. 
    
\begin{figure}
\begin{minipage}[t]{0.45\textwidth}
\justify
\def\@captype{table}
\caption{Subjective evaluation results}
\begin{tabular}{rccc}
\toprule
&\multicolumn{2}{c}{Mean rating ($N$=10)} \\
\midrule
Variable&\small{Base.(
    \begin{math}
       \sigma 
    \end{math}
)}&\small{Exp.(
    \begin{math}
       \sigma 
    \end{math}
)}&$p$\\ 
\midrule
Originality	&5.1(0.99)	&5.4(1.43)	&0.591\\
Relevancy	&5.5(1.51)	&6.1(0.99)	&0.217\\
Explainability	&5.4(1.65)	&5.9(1.45)	&0.427\\
\bottomrule
\end{tabular}
\end{minipage}
\begin{minipage}[t]{0.45\textwidth}
\begin{center}
\def\@captype{table}

\caption{Objective evaluation results}
\begin{tabular}{rccc}
\toprule
&\multicolumn{2}{c}{Mean rating ($N$=2)} \\
\midrule
Variable&\small{Base.(
    \begin{math}
       \sigma 
    \end{math}
)}&\small{Exp.(
    \begin{math}
       \sigma 
    \end{math}
)}&$p$\\ 
\midrule
Originality	&4.55(1.32)	&3.95(1.72)	&0.394\\
Relevancy	&4.95(0.64)	&4.90(0.96)	&0.893\\
Imageability	&4.45(1.30)	&4.35(0.97)	&0.848\\
\midrule
Duration(min.)	&9.02(3.76)	&8.68(4.47)	&0.888\\
W. Count in WP	&8.7(5.27)	&12.2(6.00)	&0.183\\
\bottomrule
&&&\\
\end{tabular}
\end{center}
\end{minipage}
\end{figure}

\begin{table}[h]
    \begin{center}
        \caption{Design concept phrases generated by participants}
        \begin{tabular}{cclrrrrc} \toprule
            \small{P. ID}&\small{Tool}&$w_1+w_2$&$WC$&$Freq(w_1)$&$Freq(w_2)$&mean$Freq$&$cosSim(w_1,w_2)$\\
            \midrule
            1-A	&Exp.		&``cognizant inclusive''&23	&1.02	&19.72	&10.37&0.105\\
            2-A	&Exp.		&``sustainable renewable'' &5 &97.59	&35.89	&66.74	&0.572\\
            3-A	&Exp.		&``honest continuous''	&22 &22.25 &30.06	&26.15&0.123\\
            4-A	&Exp.		&``futuristic modern'' &10 &2.22 &108.15	&55.19&0.392\\
            5-A	&Exp.		&``august renewable''	&10 &2.08	&35.89	&18.99&0.021\\
            7-B	&Exp.		&``economical efficient''	&6 &6.32	&56.97	&31.64&0.551\\
            8-B	&Exp.		&``affordable neutral'' &13 &42.35	&12.40	&27.38&0.068\\
            9-B	&Exp.		&``modular disposable'' &10 &7.16	&4.27	&5.71&0.162\\
            10-B	&Exp.		&``empathy transcendent'' &19	&null	&1.45	&1.45&0.240\\
            11-B	&Exp.		&``utilitarian comfortable'' &11 &1.13	&40.05	&20.59&0.235\\\hline
            7-A	&Base.		&``efficient functional'' &4 &56.97	&33.92	&45.45&0.382\\
            8-A	&Base.		&``good-natured safeness'' &8 &null	&null	&null&null\\
            9-A	&Base.		&``adventurous lively'' &5 &3.82	&10.20	&7.01&0.284\\
            10-A	&Base.		&``sustained delightful'' &7 &8.58	&6.24	&7.41&0.047\\
            11-A	&Base.		&``empathetic minimal'' &9  &0.98	&18.12	&9.55&0.055\\
            1-B	&Base.		&``protean companionable''	 &20 &0.17	&0.09	&0.13&0.063\\
            2-B	&Base.		&``affordable seamless'' &6 &42.35	&6.16	&24.26&0.185\\
            3-B	&Base.		&``insensible trustful'' &16	&0.20	&0.16	&0.18&0.200\\
            4-B	&Base.		&``compact friendly''	 &5 &12.77	&43.70	&28.24&0.121\\
            5-B	&Base.		&``nimble aid'' &8	&1.36	&null	&1.36&0.007\\
            \bottomrule
        \end{tabular}
    \end{center}
\end{table}

\begin{figure}
\centering
\begin{subfigure}{.5\textwidth}
  \centering
  \includegraphics[width=1\linewidth]{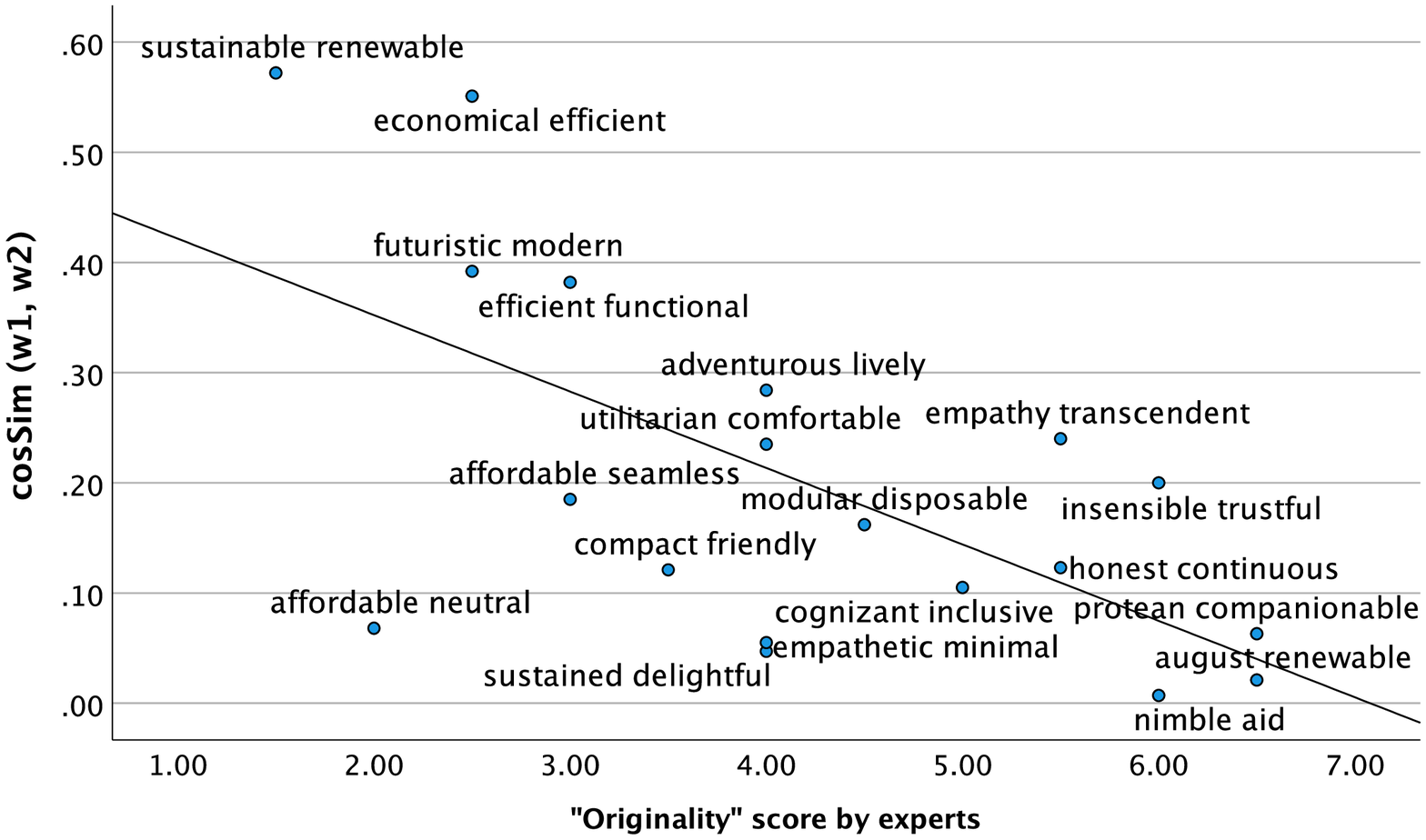}
  \caption{}
  \label{fig:sub1}
\end{subfigure}%
\begin{subfigure}{.5\textwidth}
  \centering
  \includegraphics[width=1\linewidth]{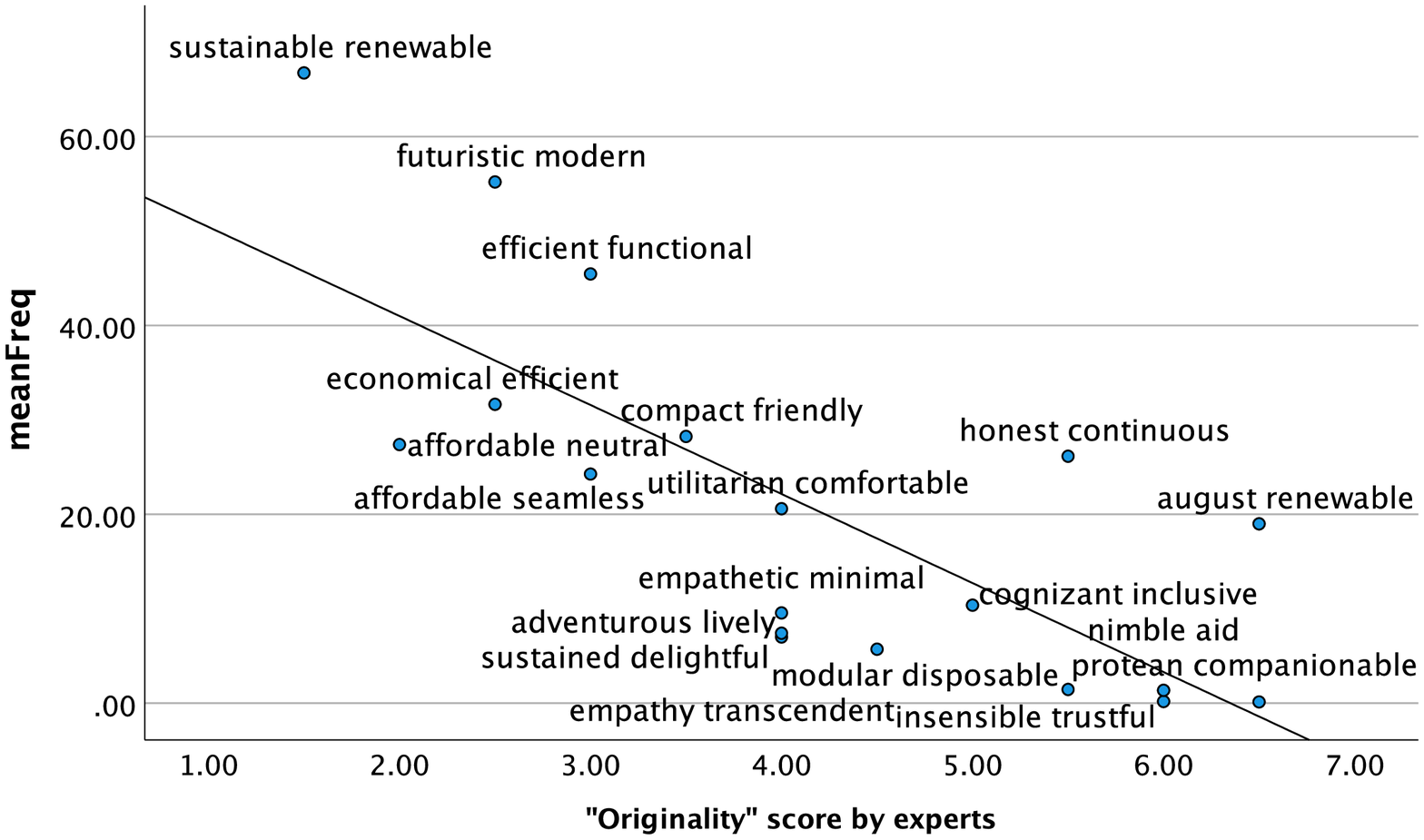}
  \caption{}
  \label{fig:sub2}
\end{subfigure}
\caption{(a) Pearson correlations between cosine similarity $(w_1, w_2)$ and experts' evaluation on ``originality.'' (b) Pearson correlations between mean mean-frequency $(w_1, w_2)$ and experts' evaluation on ``originality.''}
\label{fig:test}
\end{figure}

\begin{figure}[h]
  \centering
  \includegraphics[width=0.9\linewidth]{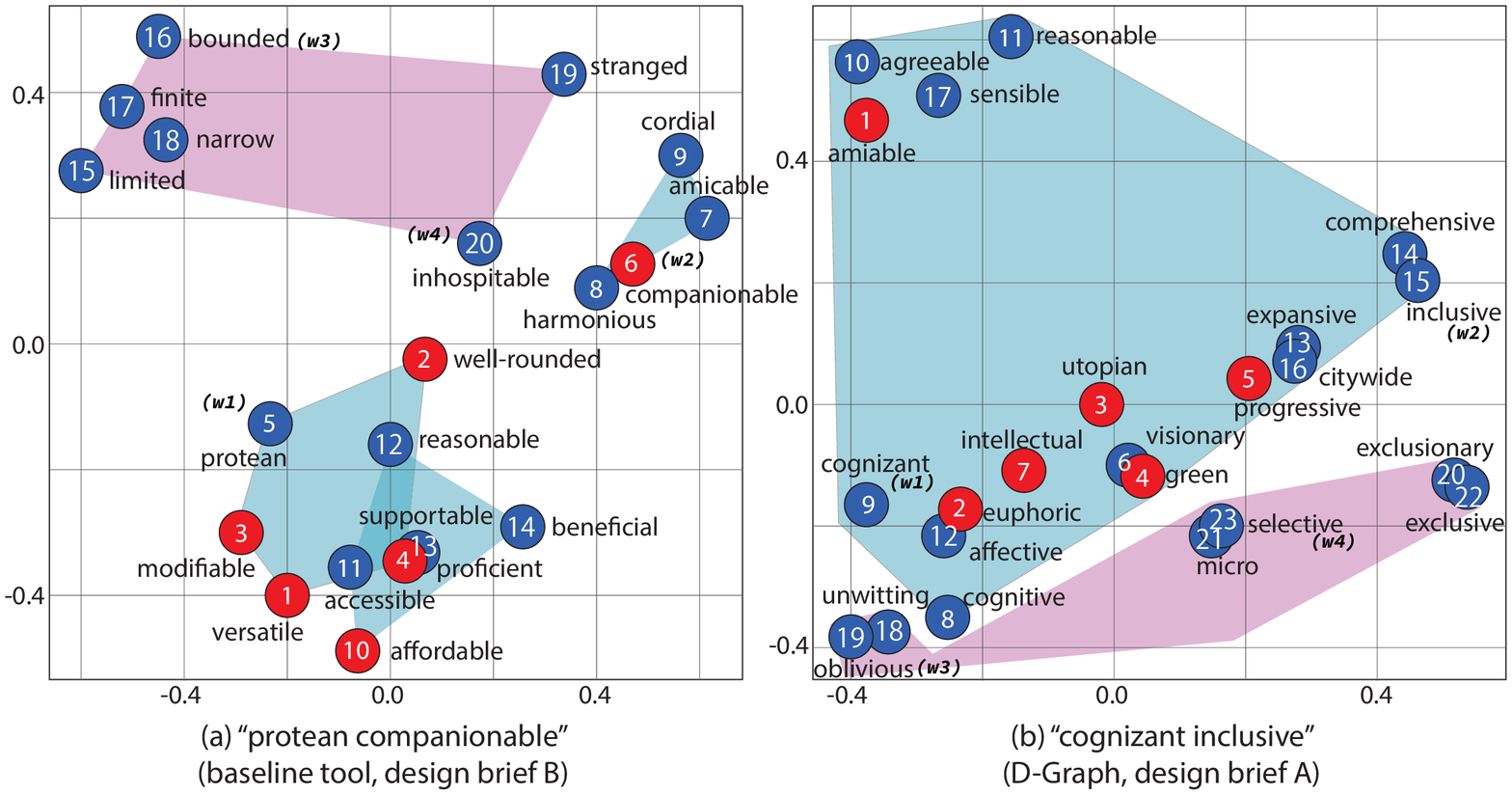}
  \caption{}
  \includegraphics[width=0.9\linewidth]{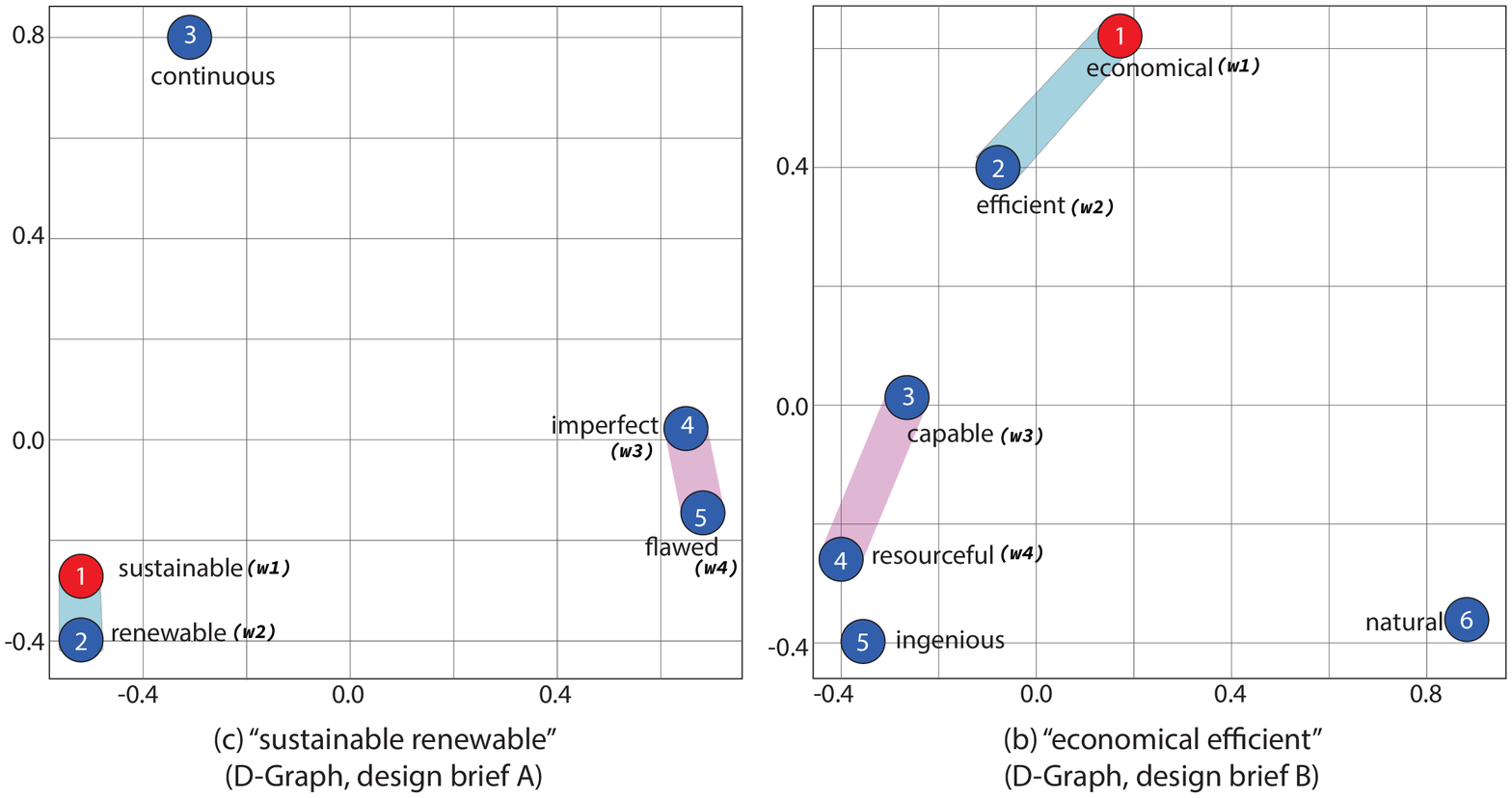}
  \caption{Sequence for word exploration in semantic space. Words were mapped according to dimensionality reduction by principal components analysis (PCA) based on ConceptNet Numberbatch word embedding. Blue areas show searches for DCPs for $w_1$ and $w_2$, and pink areas show searches for antonyms for $w_3$ and $w_4$. Red circles are users' own queries.  Blue circles are users' clicks on words.}
  \Description{}
\end{figure}

\subsection{Qualitative results}
\subsubsection{Words on character space and explanations}
All the DCPs, $w_3$ and $w_4$ on the CS, participants' ``explainability'' scores, as well as authors' observation on whether or not the explanation contains comparative descriptions using the concepts of $w_3$ and/or $w_4$ are shown on Table 6.  Only the case 9-A:``adventurous lively'' and 5-B:``nimble aid'' was explained in the way we had hoped the participants would perform. 9-A participant explains his concept as `The design concept phrase of ``adventurous lively`` speaks to the want we all have within to go out and discover new things around us with the ones we care about in a communal environment \textit{while the other concepts within the chart relate to the being more stationary and in a mellowed out environment for either self reflection or simple relaxation.}'  5-B participant explains ```Nimble Aid'' represents a combination of the agility required in the brief with overall intention of helping, \textit{whereas other phrases focus on economic or level of determination}.' 

\subsubsection{Utilization and the search algorithms}
As described in the section 5.1.3, the participant on case 1-A, who created ``cognizant inclusive'', had chosen $w_2$ from the word pool, so he did not utilize the \small{SEARCH\_FOR\_RELATED\_WORDS} \normalsize(Algorithm 1). Yet, he was able to pick two words that are distant enough to get a high ``originality'' score by the expert. He actually set the $w_3$ and $w_4$ with the words from the playground, that were the output according to the $w_1$ and $w_2$ using \small{SEARCH\_FOR\_ANTONYMS} \normalsize (Algorithm 2). This was what we had assumed how users would use the D-Graph and incorporated in the instructions to the participant. However, our video analysis unveiled there were only two cases (4-A and 5-A) that utilized the former algorithm and three cases (1-A, 4-A and 5-A) that utilized the latter algorithm to explore the words.

\subsubsection{Summary of qualitative analysis}
The analysis think-aloud protocol and screen recording uncovered variety of insights for future development and search strategies.  Below are summary of insights we would like to highlight for future development, categorized in three-fold, addressing usability issues, accommodating user's mental model, and reflections on their own DCPs, as well as on D-Graph tool.

    {\bfseries Usability issues:} First, many pointed out issues in transparency of the process and system status especially from those who were aware of the difference in phased search algorithms and tried to utilize them (1-A, 4-A, 5-A participants). It was unclear which of the two search algorithms, \small{SEARCH\_FOR\_RELATED\_WORDS} or  \small{SEARCH\_FOR\_ANTONYMS}, was running especially when users want to change some words on the axes on the CS. The weakness of the ConceptNet in showing clear antonyms also contributed to this issue. 1-A participant suggested we iterate the color system to make distinction between different search mode. In contrast to the antonyms, some participants (10-A, 2-A, 8-A) seemed to be overwhelmed by the vast number of words exploded on the D-Graph, making them difficult to recognize all words there. However, the opinions varied depending on individual comfort level.  The familiar list of Merriam Webster thesaurus gave these users more peace of mind, where as some participants (5-A, 7-A, 11-A) were excited about the novelty of UI, stating \textit{``this is more interesting than the first tool. It's easy to use as all the lines are branches showing the relationship between different words''} (7-B). Another major feedback was on the restriction of D-Graph, which does not accept non-adjective queries. (1-A, 5-A, 10-A).  In fact, no one complained about the output DCP being adjective-adjective, but the search should not block non-adjective query.  1-A participant actually has to look up an adjective form of ``utopia'' on another online dictionary before putting it into the D-Graph. Lastly, many wished they could swap the words on the CS axes.  It was, again, restricted due to the search strategy for two different search algorithm until they set all words.  Even after setting all words, swapping words between axes were not allowed. 3-B participant said, \textit{``I need to have my own logic in making a character space. Even if I find a new word, I could not shuffle, which bothers me.''}
    
    {\bfseries Accommodating users' mental model:} 
    We acknowledge that there are different mental models of the users the tool need to accommodate. Some of the participants had clear ideas of what a good DCP (some people called them `key words') consist of, and had opinions on how they achieve it. 10-A/B participants incisively pointed out, \textit{``this is good key words because two words are serving for different aspects of the design brief.''}  As a strategy, 1-A participants told us his one technique, described as \textit{``I usually use a simple query word plus `definition' on google search, that gives me a bunch of list that contains a couple of complicated words, then I put those uncommon words on thesaurus.com to look up more nuanced options.''} This participant actually performed the same way on  in creating ``definition'' on Fig. 6(a).  After exploring some fairly uncommon words, such as ``amicable'', ``cordial'', and ``harmonious'', he restarted the search with ``affordable'', mentioning \textit{``now I want to come back to a basic one.''}  Apparently he knew when to exploit and when to explore.  As such, our tool needs to accommodate the mental model they already formed, as well as that of some novice users.  In the section 5.1.3, we highlighted the long intervals participant 2-A had to take between searches to look up word definition, which by the way lots of participants appreciated, and ended up setting the first two words he found on the CS. Having seen that, creating a search strategy is one problem, but making the participants aware of the strategy is another problem we would need to address.
    
    {\bfseries Participants' reflection of the DCPs and D-Graph tool:} While the participants were asked to rate their self-reported quality of the DCP in ``originality'', ``relevancy'' and ``explainability'' they had mixed reflections on what they have created. 1-A participant initially mentioned the idea of forming a DCP did not really make sense because it was just a combination of two adjectives together after all. But after he finished his task, he stated. \textit{``I never thought up these combinations, but (when presented) it really can be something.''} 11-A participant pointed out they would need to think more with visuals, which we did not incorporate into the study this time. They noted, \textit{``thinking about concept only by words are still not enough.  I need some visual support although overall it is very fun tool to play with to see all related words.''} We did not measure engaging factors on the tool because it was outside of our scope on this study, however, it should be noted some participants expressed it was fun to interact with.  Our rational on the graphical representation was rather for a functional purpose, making it easier for the users to go back and forth in different granularity in the semantic space. We did observed some participants, like 2-A went back to the previous nodes several times to examine the direction for further explorations.
    
\begin{table}[h]
    \begin{center}
        \caption{Words on CS, explainability rating by participants, Contrasting explanation (Con.exp.)}
        \begin{tabular}{cclllcc} \toprule
            \small{P. ID}	&\small{Tool}	&$DCP(w_1 + w_2)$	&$w_3$	&$w_4$	&\small{$Explainability$}	&\small{Con. exp.}\\
            \midrule
            1-A	&Exp.	&``cognizant inclusive''	&``oblivious''	&``selective''	&7	&No\\
            2-A	&Exp.	&``sustainable renewable''	&``imperfect''	&``flawed''	&7	&No\\
            3-A	&Exp.	&``honest continuous''	&``attached''	&``instant''	&7	&No\\
            4-A	&Exp.	&``futuristic modern''	&``ordinary''	&``elite''	&3	&No\\
            5-A	&Exp.	&``august renewable''	&``civil''	&``admirable''	&4	&No\\
            7-B	&Exp.	&``economical efficient''	&``capable''	&``resourceful''	&7	&No\\
            8-B	&Exp.	&``affordable neutral''	&``comfortable''	&``independent''	&6	&No\\
            9-B	&Exp.	&``modular disposable''	&``secure''	&``easy''	&5	&No\\
            10-B	&Exp.	&``empathy transcendent''	&``conscious''	&``trustworthy''	&6	&No\\
            11-B	&Exp.	&``utilitarian comfortable''	&``economic''	&``intelligent''	&7	&No\\\hline
            7-A	&Base.	&``efficient functional''	&``dynamic''	&``kinetic''	&3	&No\\
            8-A	&Base.	&``good-natured safeness''	&``comfortable''	&``practical''	&5	&No\\
            9-A	&Base.	&``adventurous lively''	&``serene''	&``calm''	&3	&Yes\\
            10-A	&Base.	&``sustained delightful''	&``authentic''	&``modest''	&6	&No\\
            11-A	&Base.	&``empathetic minimal''	&``modular''	&``sustainable''	&7	&No\\
            1-B	&Base.	&``protean companionable''	&``bounded''	&``inhospitable''	&7	&No\\
            2-B	&Base.	&``affordable seamless''	&``special''	&``discount''	&5	&No\\
            3-B	&Base.	&``insensible trustful''	&``active''	&``regular''	&7	&No\\
            4-B	&Base.	&``compact friendly''	&``streamlined''	&``dreaming''	&4	&No\\
            5-B	&Base.	&``nimble aid''	&``resolute''	&``provident''	&7	&Yes\\
            
            \bottomrule
        \end{tabular}
    \end{center}
\end{table}


\section{Discussion and Future Work}
Our experiment did not demonstrate an advantage of the D-Graph in creating an original DCP for automotive design concept creation task in comparison with the baseline tool.  However, we observed a clear correlations between the expert's objective evaluations on DCPs and the computational linguistic measurements using the ConceptNet knowledge graph, the ConceptNet Numberbatch word embeddings and the enTenTen15 English web corpus. Our strategy was aligned with our goal, but the strategy could not be effectively implemented.  We closely examined how our participants interacted with the tool and words.  One notable finding was that the high-performers interacted not just with more words, but they also interacted with nuanced words in a semantic space.  Their strategy was rather a convergence rather than a divergence when examining words to choose from while most inquiries in creative support tools mainly focused on divergent thinking \cite{silvia2013verbal}. Our approach may be a bit confusing because on one hand we aim to maintain a distance between two concepts represented by $w_1$ and $w_2$, while on the other hand we try to support users to find nuanced words once the target semantic is set.  In the future study we will add more clarity on what strategy we want to help the users to follow, while still encouraging a sense of creative freedom which designers definitely want to maintain.  One intention is to implement more automated function we did not implemented at this time round.  For instance, extracting words from the design brief and convert non-adjective words to adjective can be automated, and would lower the initial barrier of explorations especially for inexperienced designers. Automated query generations would trigger variety of candidate at the beginning of the search.  Also, the presentation of the candidates word can be more intentional having observed it was not easy for some users to avoid cliché words.  Rather than presenting words on 3D graph, it is worth trying an organized list of words that are ranked according to the computational linguistic metrics proven to be correlate with the quality of DCPs.  We could further automate the process of concatenating two adjectives in a way they maintain certain distance we discussed in this paper. While it is still unclear what type of adjectives works better to express a product aesthetics we aim to communicate, we may need to further classify adjectives in different supersense\cite{tsvetkov2014augmenting}, witch is known to predict more of an abstract meanings of a word. Finally, we should be investigating more of fun and engaging factors\cite{cherry2014quantifying}, that we did not measure in this study, while creativity definitely need to be activated by human cognition.


\section{Conclusion}
We created an AI-assisted interactive tool, D-Graph, aiming to help designers exploring the nuanced concepts optimizing creative outcomes in verbalizing design concept during an early stage of an automotive design project. We integrated two language-based methodologies to attack the problem: 1) utilizing compound adjectives to express aesthetic design concept, 2) search and filter algorithms utilizing a ConceptNet knowledge graph, word-embedding-based language model, and corpus query system. We also implemented a dynamic 3D graph visualization for intuitive interactions. Our experiment with 10 student participants did not indicate a significant advantage with our system in comparison with the conventional online thesaurus. However, we confirmed the metrics we used, the relative word frequency and the cosine similarity between the compound words correlates with the experts objective evaluation on ``originality'' of the DCPs. Our qualitative analysis found several important aspects in how users interact with words in lexicosemantic space when searching for nuanced words to create a distinguished design concept. As for our reflections on potential social impacts our study would make, we definitely acknowledge that our system is a recommender system, whose performance may be biased depending on the algorithm and the data set from which it is trained.  While our study involves aesthetics, which is deeply rooted in a notion of the society, we will contemplate the consequences of the technology being implemented and disseminated. It is our responsibility that we research and develop these technologies in the way humanity is enhanced instead of perished.


\section{Acknowledgement}
We appreciate the support of the Academy of Art University in San Francisco, CA for referring the research participants.

\bibliographystyle{ACM-Reference-Format}
\bibliography{Sano_bib_10.2.2021}
\end{document}